\documentclass[twoside]{dis04}
\def\runauthor{Marek Ta\v{s}evsk\'{y}}
\def\shorttitle{Quark and gluon jet fragmentation functions}
\newcommand{\mr}{\mathrm}
\newcommand{\ee}{e$^+$e$^-$}

\newcommand{\qq}{${\mr q\bar{q}}$}
\newcommand{\qqg}{${\mr q\bar{q}}$g}

\newcommand{\xe}{$x_{\mr E}$}
\newcommand{\Qj}{$Q_{\mr jet}$}
\newcommand{\ff}{fragmentation function}
\newcommand{\ffs}{fragmentation functions}
\newcommand{\CLEO}{CLEO Coll.}
\newcommand{\OP}{OPAL Coll.}
\newcommand{\DE}{DELPHI Coll.}
\newcommand{\AL}{ALEPH Coll.}

\newcommand{\PRL}{Phys. Rev. Lett.}
\newcommand{\PL}{Phys. Lett.}

\newcommand{\PRV}{Phys. Rev.}
\newcommand{\EPJ}{Eur. Phys. J.}
\newcommand{\ZP}{Z. Phys.}
\newcommand{\NP}{Nucl. Phys.}

\newcommand{\al}{{\it et al.}}

\begin{document}

\title{Scaling violations of quark and gluon jet fragmentation functions
in e$^+$e$^-$ annihilations}

\author{Marek Ta\v{s}evsk\'{y}}

\address{Physics Department of the Antwerp University\\
Universiteitsplein 1, B-2610 Antwerp, Belgium}

\maketitle

\abstracts{Flavour inclusive, udsc and b \ffs\ in biased and unbiased jets  
and gluon \ffs\ in biased jets are measured in e$^+$e$^-$ annihilations using 
OPAL data collected at energies $\sqrt s =$ 91.2 and 183--209 GeV.}

\section{Introduction}
Hadron production in high energy collisions can be described by parton showers,
followed by the formation of
hadrons which cannot be described perturbatively. Gluon emission, the dominant
process in parton showers, is proportional to the colour factor associated
with the coupling of the emitted gluon to the emitter. These colour factors
are $C_A=3$ when the emitter is a gluon and $C_F=4/3$ when it is a quark.
Consequently, the multiplicity of soft gluons from a gluon source is
(asymptotically) 9/4 times higher than from a quark source \cite{Brodsky}.
The inequality between $C_A$ and $C_F$ plays a key role in the explanation of
the observed differences between quark and gluon jets: compared to quark jets,
gluon jets are observed to have larger widths \cite{qgdif}, higher
multiplicities \cite{qgdif,subjet}, softer \ffs\ \cite{qgdif,gincl3,Master},
and stronger scaling violations of the \ffs\ \cite{Master}. 

The \ff, $D_a^h(x,Q^2)$, is defined as the probability that parton $a$, which
is produced at short distance, of order $1/Q$, fragments into hadron, $h$,
carrying the fraction $x$ of the momentum of $a$. In this study, the momentum
fraction is defined as \xe\ $=E_h/E_{\mr jet}$, where $E_h$ is the energy of
the hadron $h$ and $E_{\mr jet}$ is the energy of the jet to which it is
assigned. 

Three-jet \qqg\ events are selected by applying a jet finder.
Different jet finders result in different
assignments of particles to jets: thus jets defined using a jet finding
algorithm are called {\it biased}. In contrast, quark and gluon jets used in
theoretical calculations are usually defined as inclusive hemispheres of 
back-to-back \qq\ and gg final states, respectively. The hemisphere
definition yields a so-called {\it unbiased} jet because the jet properties do
not depend on the choice of a jet finder. Measurements of unbiased quark jets
have been performed at many scales \cite{loweren,B-decay,flavourff}. Direct
measurements of unbiased gluon jets are so far available only from the CLEO
\cite{CLEO} and OPAL \cite{gincl3,gincl12} experiments, however. 
Recently, the OPAL experiment has measured properties of unbiased gluon jets
indirectly \cite{indgl1,indgl2}. 

In our study, we present measurements of quark
and gluon jet \ffs\ at $\sqrt s =$ 91.2~GeV and $\sqrt s =$ 183--209~GeV.
The data were collected with the OPAL detector at the LEP \ee\ collider at
CERN. We measured seven types of \ffs: those from biased as well as unbiased 
flavour inclusive, udsc and b jets, and from biased gluon jets. While the two 
types of flavour inclusive jets have been measured many times, data on the 
other types of \ffs\ are still rather scarce. 

\section{Analysis procedure}\label{procedure}
The selection of hadronic Z and Z$^*/\gamma^*$ events and any other details
of this analysis are described in \cite{marek}. In the inclusive hadronic event 
samples, we use the unbiased jet definition where the jets are defined by 
particles in hemispheres of the \qq\ system. In the three-jet samples, we apply 
a jet algorithm and thus work with biased jets. Three jet algorithms are used: 
the Durham \cite{Durham}, Cambridge \cite{Cambridge} and cone \cite{cone} 
algorithms. The jet algorithm is forced to resolve three jets per event. 
The jet energies and momenta are then recalculated by imposing overall 
energy-momentum conservation with planar massless kinematics, using the jet 
directions found by the jet algorithm. 

The measured \ff\ is defined here as the total number of
charged particles, $N_p$, in bins of \xe\ and scale $Q$ normalized to the
number of jets, $N_{\mr jet}(Q)$, in the bin of $Q$:
\begin{equation}
\frac{1}{N_{\mr jet}(Q)}\frac{{\mathrm d}N_p(x_{\mr E},Q)}
{{\mathrm d}x_{\mr E}}
\label{ffdef}
\end{equation}

To measure the scale dependence, it is necessary to specify a scale
relevant to the process under study. For inclusive hadronic events, the scale
is $\sqrt s$. For jets in three-jet events, neither $\sqrt s$ nor
$E_{\mr jet}$ is considered to be an appropriate choice of the scale
\cite{Qjet}. QCD coherence suggests \cite{QCDBas} that the event topology
should also be taken into account. Similarly to previous studies, the transverse 
momentum-like scale, $Q_{\mr jet} = E_{\mr jet}\sin(\vartheta/2)$, is used
where $\vartheta$ is the angle between the jet with $E_{\mr jet}$ and the 
closest other jet.

In this analysis, three methods are used to identify quark and gluon jets: the 
b-tag and the energy-ordering methods in biased three-jet events, and the 
hemisphere method in unbiased inclusive hadronic events. In addition, b
tagging is used to separate udsc and b quark jets from each other, both for
the biased and unbiased jet samples. The energy-ordering method only allows 
flavour inclusive quark jets to be distinguished from gluon jets. 

All three methods are in detail described in \cite{marek}. 
In three-jet events, any of the three jets is used to extract the \ffs. The 
purity of the b-tag (gluon) jet 
sample in LEP1 data is estimated to be 90\% (84\%) and the efficiencies are 23\% 
(40\%). The corresponding purities in LEP2 data are 60\% (80\%) and efficiencies 
27\% (45\%). The purity of the b-tag hemispheres in LEP1 data is estimated to be
99.7\%, while in LEP2 data it is 75\%. 

After subtracting the remaining background from LEP2 data using MC estimates,
the data and MC distributions at detector level are multiplied by corresponding
inverse matrices to get the distributions at level of pure quarks and gluons.
As a last step, the data are corrected for effects of limited detector acceptance
and resolution using correction factors obtained from MC events.

\section{Results}\label{results}
The measured \ffs\ are presented with emphasis on the scale dependence (the \xe\ 
dependence and other comparisons are shown in \cite{marek}) and are compared with
previous measurements as well as with theoretical NLO predictions. The term scale
in the following figures stands for \Qj\ in case of biased jets and 
$\sqrt {s}/2$ in case of unbiased jets. The published unbiased jet results and 
the NLO predictions are scaled by $\frac{1}{2}$ since they refer to the entire 
event, thus to two jets. 

In Figs.~1a)--d) the results for the udsc, b, gluon and flavour inclusive jet 
\ffs\ are presented. The LEP1 unbiased jet data correspond to $\sqrt s =$ 
91.2~GeV. The LEP2 unbiased b jets are measured in $\sqrt s$ range of 
183--209~GeV, while the unbiased udsc and flavour inclusive jets are measured in 
three $\sqrt s$ intervals: 183--189, 192--202 and 204--209~GeV. 
The quark biased jet data from LEP1 cover the region $Q_{\mr jet}=$ 4--42~GeV,
while those from LEP2 cover the region $Q_{\mr jet}=$ 30--105~GeV.
The results from the region $0.01<x_{\mr E}<0.03$ are not shown but they are
presented in \cite{marek}. As also discussed in \cite{marek}, the results are 
found to be consistent with previous measurements. 
The udsc jet results above 45.6~GeV, the gluon jet results above 30~GeV (except 
for the ${\mr g_{incl}}$ jets, see \cite{gincl3}), and the b jet results at all 
scales except 45.6~GeV represent the first measurements.

The data are compared to three theoretical predictions: Kniehl, Kramer and 
P\"{o}tter (KKP) \cite{KKP}, Kretzer (Kr) \cite{Kretzer} and Bourhis, Fontannaz,
Guillet and Werlen (BFGW) \cite{BFGW}. 
For the udsc jet \ff\ (Fig.~1a)), all three predictions
give a good description in the entire measured phase space, except for
the lowest \xe\ bin where the KKP calculations overestimate the data, and the
highest \xe\ bin where the data are underestimated by the Kr and BFGW
calculations. 

The situation is rather different for the b and gluon jet \ffs\
(Figs.~1b) and 1c)) where the description of the data by the
NLO predictions is worse and where there are significant differences between
individual NLO results, the latter being expected due to
differences in the calculations as discussed in \cite{marek}.
In Fig.~1b) the KKP prediction is deficient with respect to the data
for $x_{\mr E}>0.12$. 
In this context, it is important to note that B hadron decay products
are indirectly included in theory predictions since they are present in the
data sets to which the fits were made. 

For the gluon jet \ffs, the two alternative methods of identifying gluon jets
described above are examined, see Fig.~1c).
The \Qj\ binning is not the same for the two methods because of their different
regions of applicability. 
A satisfactory correspondence between the b-tag and energy-ordering methods is
found in the entire scale range accessible. The data tend to show larger
scaling violations than predicted by any of the calculations. 

The results for the flavour inclusive jet \ffs, presented in Fig.~1d), are
seen to be consistent with published unbiased jet data from lower energy \ee\ 
experiments (TASSO, MARK II, TPC and AMY) \cite{loweren} and previous OPAL 
results \cite{OP133,OP161,OP172}. 
The data are also compared to the NLO predictions of KKP, Kr and BFGW which
all give a reasonable description of the data in the region of 
$0.06 < x_{\mr E} < 0.60$ and over the entire scale range. 

A good correspondence found between the results from biased and unbiased
jets in all four figures suggests that \Qj\ is an
appropriate choice of scale in three-jet events with a general topology.
A similar conclusion was previously made in \cite{Master}. The MC
study presented in \cite{marek}, however, demonstrates that the
bias introduced in the gluon jet identification is not negligible for 
$x_{\mr E}>0.6$.
In each of these figures, the scaling violation seen in the data is positive
for low \xe\ and negative for high \xe. It is more pronounced in the gluon jets
than in the quark jets.

\newpage
\vspace*{0.5cm}
\begin{minipage}{30pc}
\hspace*{-0.4cm}
\epsfig{file=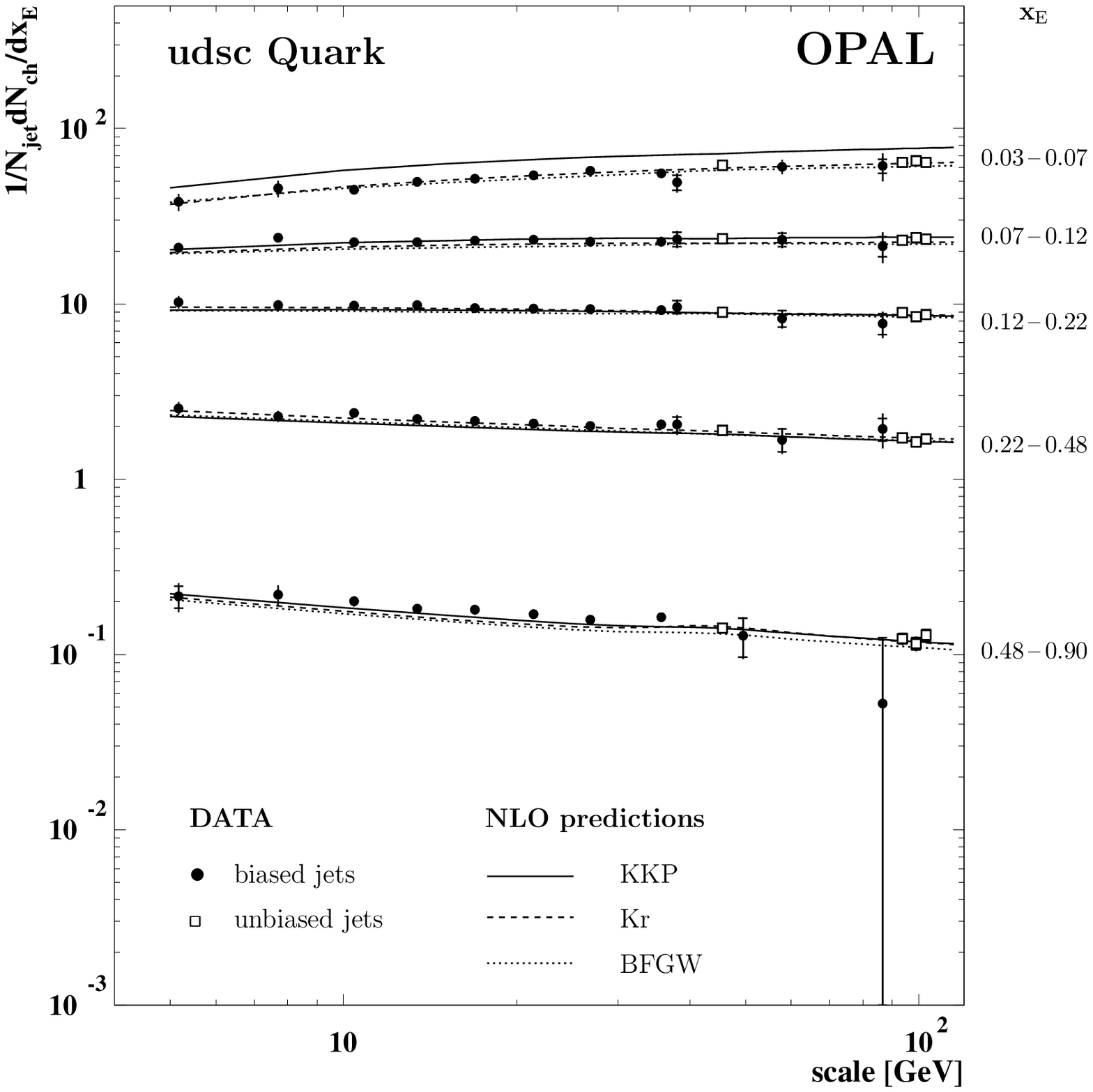,height=7cm,width=6.2cm,%
bbllx=58pt,bblly=190pt,bburx=538pt,bbury=670pt}
\put(-80,189) {\small a)}
\hspace{0.1cm}
\epsfig{file=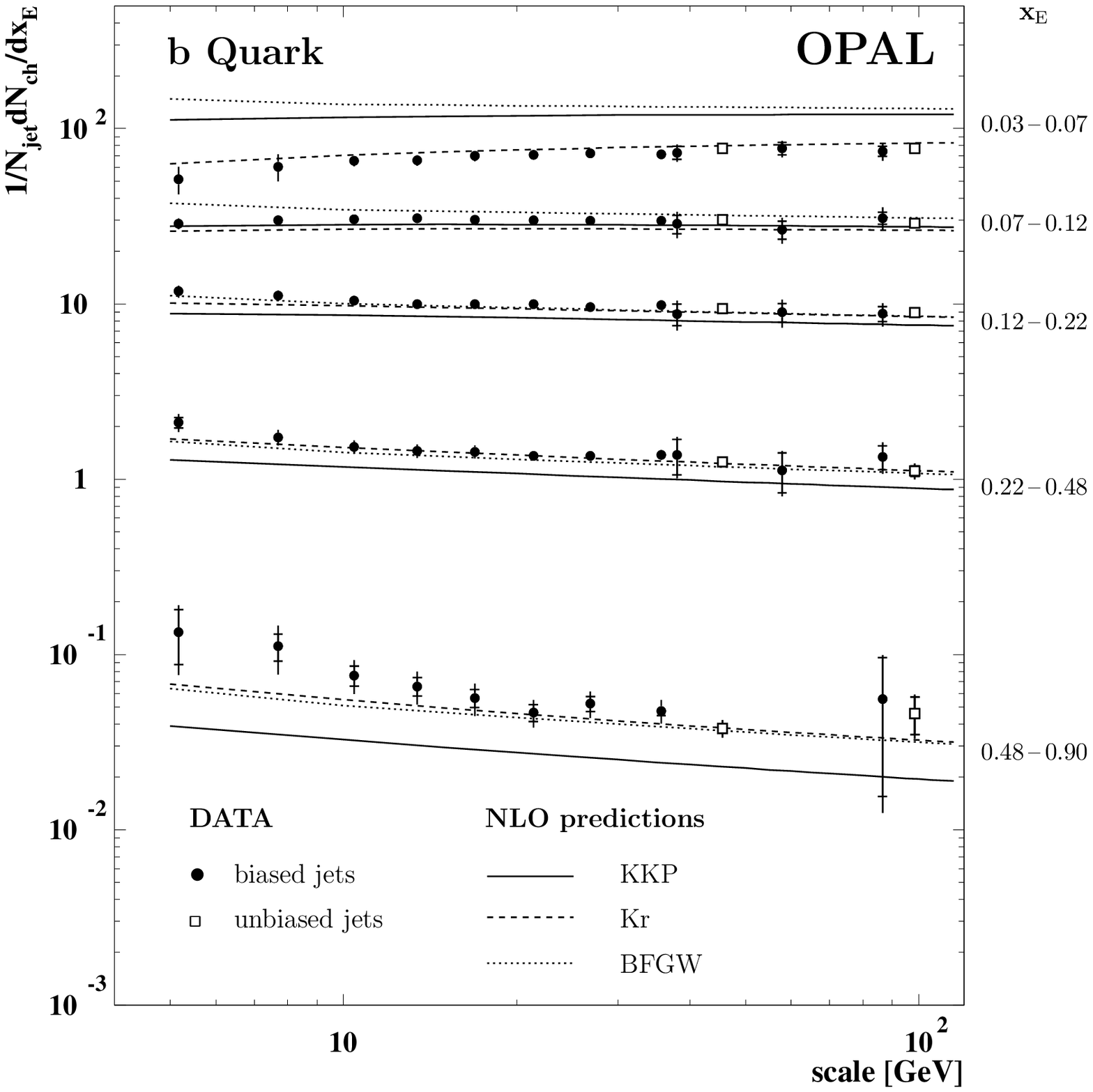,height=7cm,width=6.2cm,%
bbllx=58pt,bblly=190pt,bburx=538pt,bbury=670pt}
\put(-80,189) {\small b)}
\end{minipage}

\vspace*{0.5cm}

\begin{minipage}{30pc}
\hspace*{-0.4cm}
\epsfig{file=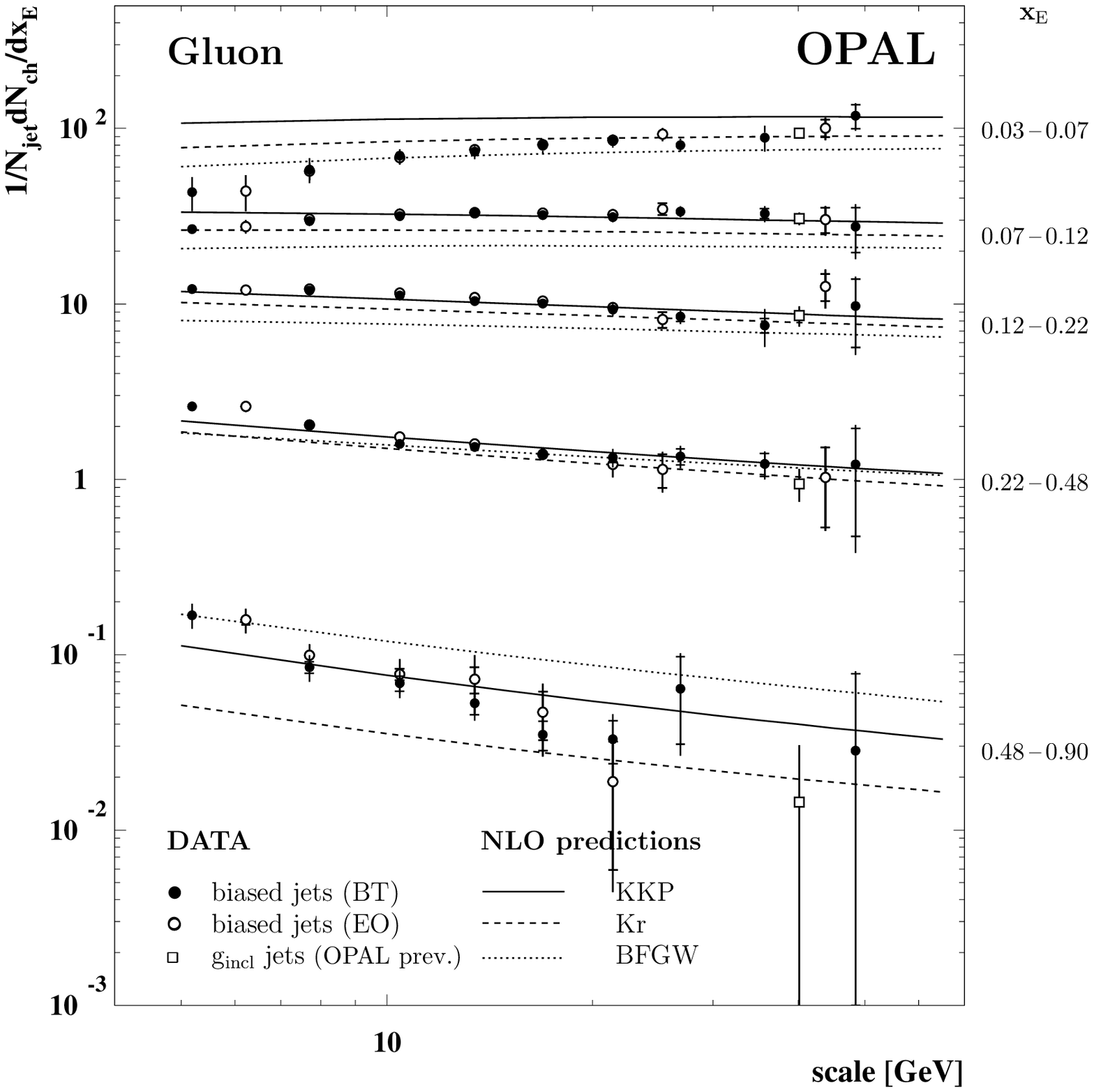,height=7cm,width=6.2cm,%
bbllx=58pt,bblly=185pt,bburx=538pt,bbury=665pt}
\put(-80,189) {\small c)}
\hspace{0.1cm}
\epsfig{file=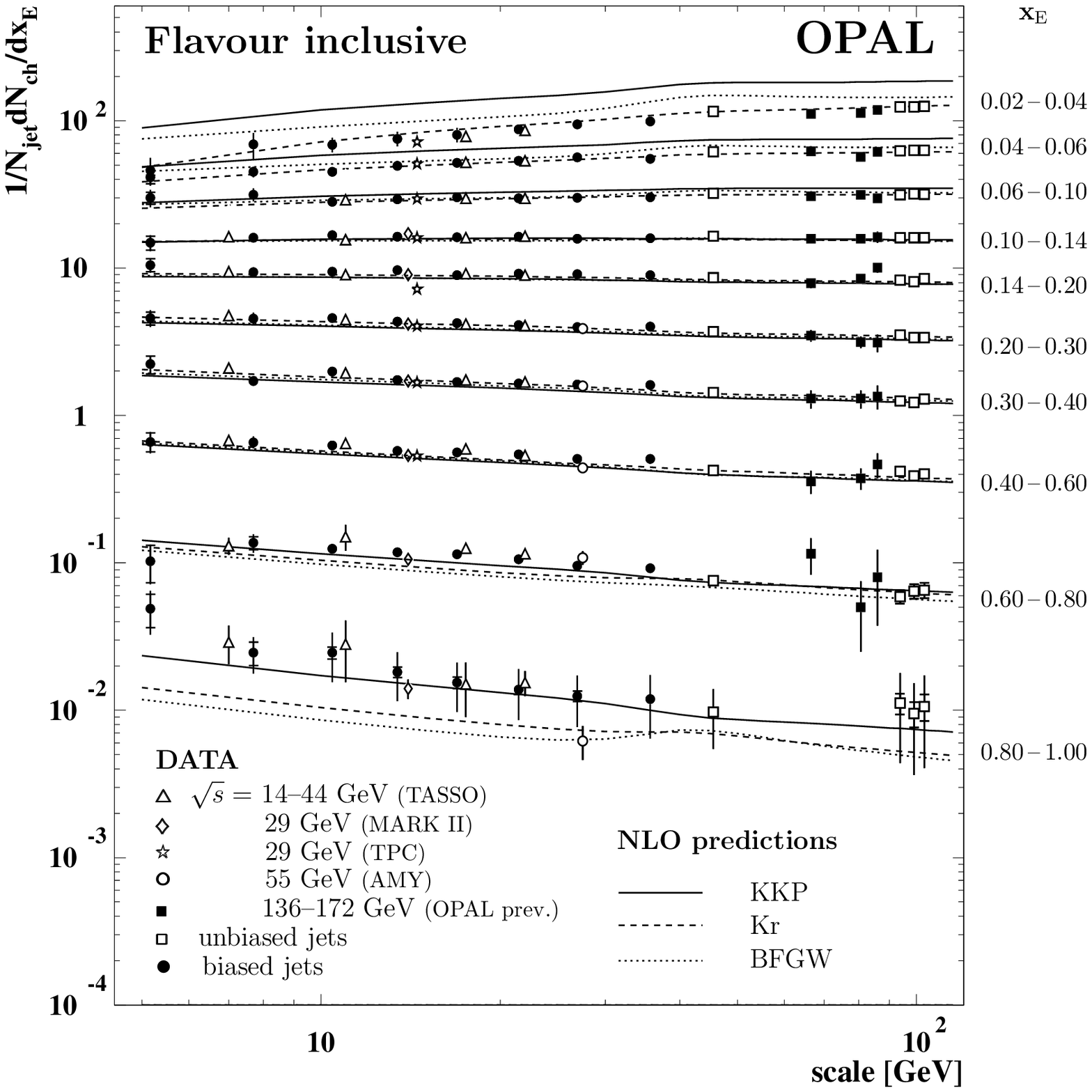,height=7cm,width=6.2cm,%
bbllx=57pt,bblly=190pt,bburx=538pt,bbury=670pt}
\put(-80,190) {\small d)}
\end{minipage}

\vspace*{0.5cm}

{\parbox[t]{30pc}
{\footnotesize Figure 1: Scale dependence of \ffs\ in different \xe\ bins. 
The scale denotes \Qj\ for the biased jets and $\sqrt s/2$ for the unbiased 
jets. The inner error bars indicate the statistical uncertainties, the total 
error bars show the statistical and systematic uncertainties added in 
quadrature. The data are compared to the NLO predictions by KKP \cite{KKP}, 
Kr \cite{Kretzer} and BFGW \cite{BFGW}. a) udsc jets, b) b jets, c) gluon
jets from b-tag (BT) method and energy-ordering (EO) method. In addition, the
published ${\mr g_{incl}}$ jets \cite{gincl3} are shown at $E_{\mr jet}=40.1$ 
GeV. d) flavour inclusive jets. In addition, the published unbiased jet data by 
TASSO, TPC, MARK II, AMY \cite{loweren} and OPAL \cite{OP133,OP161,OP172} are 
shown. }


\begin{thebibliography}{0}
%Introduction
\bibitem{Brodsky} S.J. Brodsky and J.F. Gunion, \PRL\ {\bf 37} (1976) 402;\\
K. Konishi, A. Ukawa and G. Veneziano, \PL\ {\bf B78} (1978) 243.
\bibitem{qgdif} \OP, G. Alexander \al, \PL\ {\bf B265} (1991) 462; \OP, P. Acton 
\al, \ZP\ {\bf C58} (1993) 387; \OP, R. Akers \al, \ZP\ {\bf C68} (1995) 179; 
\DE, P. Abreu \al, \ZP\ {\bf C70} (1996) 179; \AL, D. Buskulic \al, \PL\ 
{\bf B384} (1996) 353.
\bibitem{subjet}\AL, D. Buskulic \al, \PL\ {\bf B346} (1995) 389.
\bibitem{gincl3} \OP, G. Abbiendi \al, \EPJ\ {\bf C11} (1999) 217.
\bibitem{Master} \DE, P. Abreu \al, \EPJ\ {\bf C13} (2000) 573.
\bibitem{loweren}
TASSO Coll., \ZP\ {\bf C 47} (1990) 187; MARK II Coll., A. Peterson \al, \PRV\ 
{\bf D37} (1998) 1; TPC Coll., H. Aihara \al, \PRL\ {\bf 61} (1988) 1263; AMY 
Coll., Y.K. Li \al, \PRV\ {\bf D41} (1990) 2675.
\bibitem{B-decay} \DE, P. Abreu \al, \EPJ\ {\bf C5} (1998) 585.
\bibitem{flavourff} \OP, K. Ackerstaff \al, \EPJ\ {\bf C7} (1999) 369.
\bibitem{CLEO} \CLEO, M.S. Alam \al, \PRV\ {\bf D46} (1992) 4822; \CLEO, M.S. 
Alam \al, \PRV\ {\bf D56} (1997) 17.
\bibitem{gincl12} \OP, G. Alexander \al, \PL\ {\bf B388} (1996) 659; \OP, 
K. Ackerstaff \al, \EPJ\ {\bf C1} (1998) 479.
\bibitem{indgl1} \OP, G. Abbiendi \al, \EPJ\ {\bf C23} (2002) 597.
\bibitem{indgl2} \OP, G. Abbiendi \al, Phys. Rev. {\bf D69} (2004) 032002.

%Event and jet selection
\bibitem{marek} \OP, G. Abbiendi \al, submitted to \EPJ, hep-ex/0404026.
\bibitem{Durham} S. Catani \al, \PL\ {\bf B269} (1991) 432.
\bibitem{Cambridge} Yu.L. Dokshitzer \al, JHEP {\bf 9708} (1997) 001.
\bibitem{cone} UA1 Coll., G. Arnison \al, \PL\ {\bf B122} (1983) 103;
J.E. Huth \al, Ed. E.L. Berger, World Scientific,
Singapore (1990) 134; \OP, R. Akers \al, \ZP\ {\bf C63} (1994) 197.
\bibitem{Qjet} \AL, D. Buskulic \al, \ZP\ {\bf C76} (1997) 191.
\bibitem{QCDBas} Yu. Dokshitzer \al, Basics of perturbative QCD,
Editions Fronti\`{e}res (1991).

%Results
\bibitem{KKP} B.A. Kniehl, G. Kramer and B. P\"{o}tter, \NP\ {\bf B582} (2000)
514.
\bibitem{Kretzer} S. Kretzer, \PRV\ {\bf D62} (2000) 054001.
\bibitem{BFGW} L. Bourhis \al, \EPJ\ {\bf C19} (2001) 89.
\bibitem{OP133} \OP, G. Alexander \al, \ZP\ {\bf C72} (1996) 191.
\bibitem{OP161} \OP, K. Ackerstaff \al, \ZP\ {\bf C75} (1997) 193.
\bibitem{OP172} \OP, G. Abbiendi \al, \EPJ\ {\bf C16} (2000) 185.

\end{thebibliography}
\end{document}